\documentclass{ws-ijmpd}
\def\Journal#1#2#3#4{{#1} {\bf #2} (#4) #3 }

\def\RNC{\em Rivista Nuovo Cimento}

\def\NIMA{{\em Nucl. Instrum. Methods} A}

\def\PLB{{\em Phys. Lett.}  B}
\def\PRL{\em Phys. Rev. Lett.}
\def\PRD{{\em Phys. Rev.} D}

\def\GaC{\em Gravitation and Cosmology}

\def\JETPL{\em JETP Lett.}

\def\CQG{\em Class. Quantum Grav.}
\def\APJ{\em Astrophys. J.}
\def\SCI{\em Science}
\def\MPLA{{\em Mod. Phys. Lett.}  A}
\def\IJTP{\em Int. J. Theor. Phys.}
\def\NJP{\em New J. of Phys.}
\def\JHEP{\em JHEP}
\def\BWP{\em Bled Workshops in Physics}
\def\EPHJ{\em Eur.Phys.J}
\def\s{{\,\rm s}}
\def\g{{\,\rm g}}
\def\eV{\,{\rm eV}}
\def\keV{\,{\rm keV}}
\def\MeV{\,{\rm MeV}}
\def\GeV{\,{\rm GeV}}
\def\TeV{\,{\rm TeV}}
\def\sv{\left<\sigma v\right>}
\def\({\left(}
\def\){\right)}
\def\cm{{\,\rm cm}}

\def\mK{{\,\rm mK}}
\def\kpc{{\,\rm kpc}}
\def\beq{\begin{equation}}
\def\eeq{\end{equation}}
\def\bea{\begin{eqnarray}}
\def\eea{\end{eqnarray}}
\begin{document}

\markboth{M.Yu.Khlopov, A.G.Mayorov, E.Yu.Soldatov}
{Composite Dark Matter and Puzzles of Dark Matter Searches}

%
\catchline{}{}{}{}{}
%

\title{Composite Dark Matter and Puzzles of Dark Matter Searches }

\author{MAXIM YU. KHLOPOV}

\address{Virtual Institute of Astroparticle physics, APC laboratory \\
10, rue Alice Domon et L\'eonie Duquet 75205
Paris Cedex 13, France,\\ Centre for Cosmoparticle Physics "Cosmion" and \\
National Research Nuclear University (Moscow Engineering Physics Institute)\\
115409 Moscow, Russia \\
khlopov@apc.univ-paris7.fr}

\author{ANDREY G. MAYOROV}

\address{National Research Nuclear University (Moscow Engineering Physics Institute)\\
115409 Moscow, Russia\\
mayorov.a.g@gmail.com}

\author{EVGENY YU. SOLDATOV}

\address{National Research Nuclear University (Moscow Engineering Physics Institute)\\
115409 Moscow, Russia\\
Evgeny.Soldatov@cern.ch}

\maketitle

\begin{history}
\received{Day Month Year}
\revised{Day Month Year}
\comby{Managing Editor}
\end{history}

\begin{abstract}
Positive results of dark matter searches in DAMA/NaI and DAMA/LIBRA experiments, being put together with the results of other groups, can imply nontrivial particle physics solutions for cosmological dark matter. Stable particles with charge -2, bound with primordial helium in O-helium "atoms" (OHe), represent a specific
Warmer than Cold nuclear-interacting form of dark matter. Slowed down in the
terrestrial matter, OHe is elusive for direct methods of underground
Dark matter detection used in cryogenic experiments. However radiative capture of OHe by Na and I nuclei can lead to annual variations of energy
release in the interval of energy 2-5 keV in DAMA/NaI and DAMA/LIBRA
experiments.
\end{abstract}

\keywords{Elementary particles, nuclear reactions,
dark matter.}

The widely shared belief is that the dark matter, corresponding to
$25\%$ of the total cosmological density, is nonbaryonic and
consists of new stable particles. One can formulate the set of
conditions under which new particles can be considered as candidates
to dark matter (see e.g. Refs.~\refcite{book}--\refcite{Bled07} for
review and reference): they should be stable, saturate the measured
dark matter density and decouple from plasma and radiation at least
before the beginning of matter dominated stage. The easiest way to
satisfy these conditions is to involve neutral weakly interacting
particles. However it is not the only particle physics solution for
the dark matter problem. In the composite dark matter scenarios new
stable particles can have electric charge, but escape experimental
discovery, because they are hidden in atom-like states maintaining
dark matter of the modern Universe.

It offers new solutions for the
physical nature of the cosmological dark matter. The main problem
for these solutions is to suppress the abundance of positively
charged species bound with ordinary electrons, which behave as
anomalous isotopes of hydrogen or helium. This problem is
unresolvable, if the model predicts stable particles with charge -1,
as it is the case for tera-electrons \cite{Glashow,Fargion:2005xz}.
To avoid anomalous isotopes overproduction, stable particles with
charge -1 should be absent, so that stable negatively charged
particles should have charge -2 only.

Elementary particle frames for heavy stable -2 charged species are provided by:
(a) stable "antibaryons" $\bar U \bar U \bar U$ formed by anti-$U$ quark of fourth generation
\cite{Q,I,lom,Khlopov:2006dk} (b) AC-leptons \cite{Khlopov:2006dk,5,FKS}, predicted in the
extension \cite{5} of standard model, based on the approach of
almost-commutative geometry \cite{bookAC}.  (c) Technileptons and anti-technibaryons
\cite{KK} in the framework of walking
technicolor models (WTC) \cite{Sannino:2004qp}. (d) Finally, stable
charged clusters $\bar u_5 \bar u_5 \bar u_5$ of (anti)quarks $\bar
u_5$ of 5th family can follow from the approach, unifying spins and
charges \cite{Norma}.

In the asymmetric case, corresponding to excess of -2 charge
species, $X^{--}$,
their positively charged antiparticles
effectively annihilate in the early Universe. In all the models,
in which new stable species belong to non-trivial representations of electroweak SU(2) group
sphaleron transitions at high temperatures
provide the relationship between baryon asymmetry and excess of -2 charge stable species.

 After it is formed
in the Standard Big Bang Nucleosynthesis (SBBN), $^4He$ screens the
$X^{--}$ charged particles in $(^4He^{++}X^{--})$
{\it O-helium} ``atoms'' \cite{I}.

In all the forms of O-helium, $X^{--}$ behaves either as lepton or
as specific "heavy quark cluster" with strongly suppressed hadronic
interaction. Therefore O-helium interaction with matter is
determined by nuclear interaction of $He$. These neutral primordial
nuclear interacting objects contribute to the modern dark matter
density and play the role of a nontrivial form of strongly
interacting dark matter \cite{Starkman}.

Here after a brief review of main features of OHe Universe we
concentrate on its effects in underground detectors. We present
a quantitative confirmation of the earlier guess \cite{I,I2,KK2,Bled09,unesco} that
the positive results of dark matter searches in DAMA/NaI (see for
review Ref.~\refcite{Bernabei:2003za}) and DAMA/LIBRA \cite{Bernabei:2008yi}
experiments can be explained by effect of O-helium, resolving the controversy
between these data and negative results of other experimental
groups.

\section{O-helium Universe}

Following Refs.~\refcite{I}--\refcite{Khlopov:2006dk},\refcite{KK} and \refcite{I2} consider charge
asymmetric case, when excess of $X^{--}$ provides effective
suppression of positively charged species.

In the period $100\s \le t \le 300\s$  at $100 \keV\ge T \ge T_o=
I_{o}/27 \approx 60 \keV$, $^4He$ has already been formed in the
SBBN and virtually all free $X^{--}$ are trapped by $^4He$ in
O-helium ``atoms" $(^4He^{++} X^{--})$. Here the O-helium ionization
potential is\footnote{The account for charge distribution in $He$
nucleus leads to smaller value $I_o \approx 1.3 \MeV$
\cite{Pospelov}.} \beq I_{o} = Z_{x}^2 Z_{He}^2 \alpha^2 m_{He}/2
\approx 1.6 \MeV,\label{IO}\eeq where $\alpha$ is the fine structure
constant,$Z_{He}= 2$ and $Z_{x}= 2$ stands for the absolute value of
electric charge of $X^{--}$.  The size of these ``atoms" is
\cite{I,FKS} \beq R_{o} \sim 1/(Z_{x} Z_{He}\alpha m_{He}) \approx 2
\cdot 10^{-13} \cm \label{REHe} \eeq Here and further, if not
specified, we use the system of units $\hbar=c=k=1$.


Due to nuclear interactions with nuclei of
cosmic plasma, the O-helium gas is in thermal equilibrium with
plasma and radiation on the Radiation Dominance (RD) stage, while
the energy and momentum transfer from plasma is effective. The
radiation pressure acting on the plasma is then transferred to
density fluctuations of the O-helium gas and transforms them in
acoustic waves at scales up to the size of the horizon.

At temperature $T < T_{od} \approx 200 S^{2/3}_3\eV$ the energy and
momentum transfer from baryons to O-helium is not effective
\cite{I,KK} because $n_B \sv (m_p/m_o) t < 1,$ where $m_o$ is the
mass of the $OHe$ atom, $S_3= m_o/(1 \TeV)$, $m_p$ is the mass of proton, $\sigma
\approx \sigma_{o} \sim \pi R_{o}^2 \approx
10^{-25}\cm^2$ and $v = \sqrt{3T/m_p}$ is the
baryon thermal velocity. Then O-helium gas decouples from plasma. It
starts to dominate in the Universe after $t \sim 10^{12}\s$  at $T
\le T_{RM} \approx 1 \eV$ and O-helium ``atoms" play the main
dynamical role in the development of gravitational instability,
triggering the large scale structure formation. The composite nature
of O-helium determines the specifics of the corresponding dark
matter scenario, which has
 qualitative feature of a Warmer Than Cold
Dark Matter model \cite{unesco}.

Being decoupled from baryonic matter, the $OHe$ gas does not follow
the formation of baryonic astrophysical objects (stars, planets,
molecular clouds...) and forms dark matter halos of galaxies. It can
be easily seen that O-helium gas is collisionless for its number
density, saturating galactic dark matter. Taking the average density
of baryonic matter one can also find that the Galaxy as a whole is
transparent for O-helium in spite of its nuclear interaction. Only
individual baryonic objects like stars and planets are opaque for
it.

O-helium atoms can be destroyed in astrophysical processes, giving
rise to acceleration of free $X^{--}$ in the Galaxy.

If the mechanisms of $X^{--}$ acceleration are effective, the
anomalous low $Z/A$ component of $-2$ charged $X^{--}$ can be
present in cosmic rays at the level \cite{unesco,Mayorov} $X/p \sim n_{X}/n_g \sim
10^{-9}S_3^{-1},$ and be within the reach for PAMELA and AMS02
cosmic ray experiments.

In the framework of Walking Technicolor model the excess of both stable $X^{--}$
and $Y^{++}$ is possible \cite{KK2}, the latter being two-three orders of magnitude smaller, than the former.
It leads to the two-component composite dark matter scenario with the dominant OHe accompanied by a subdominant WIMP-like component of $(X^{--}Y^{++})$ bound systems. Technibaryons and technileptons can be metastable and decays of $X^{--}$ and $Y^{++}$ can provide explanation for anomalies, observed in high energy cosmic positron spectrum by PAMELA and in high energy electron spectrum by FERMI and ATIC.

 O-helium collisions in the galactic bulge can lead to excitation of O-helium. If 2S
level is excited, pair production dominates over two-photon channel
in the de-excitation by $E0$ transition and positron production with
the rate $3 \cdot 10^{42}S_3^{-2} \s^{-1}$ is not accompanied by
strong gamma signal. According to Ref.~\refcite{Finkbeiner:2007kk} this rate
of positron production for $S_3 \sim 1$ is sufficient to explain the
excess in positron annihilation line from bulge, measured by
INTEGRAL (see Ref.~\refcite{integral} for review and references). If $OHe$
levels with nonzero orbital momentum are excited, gamma lines should
be observed from transitions ($ n>m$) $E_{nm}= 1.598 \MeV (1/m^2
-1/n^2)$ (or from the similar transitions corresponding to the case
$I_o = 1.287 \MeV $) at the level of $3 \cdot 10^{-4}S_3^{-2}(\cm^2 \s
\MeV sr)^{-1}$.

\section{O-helium in the terrestrial matter}
The evident consequence of the O-helium dark matter is its
inevitable presence in the terrestrial matter, which appears opaque
to O-helium and stores all its in-falling flux.

The nuclear cross section of the O-helium
interaction with matter escapes the severe constraints
on strongly interacting dark matter particles
(SIMPs) \cite{Starkman} imposed by the XQC experiment
\cite{XQC}. Therefore, a special strategy of direct O-helium  search
is needed, as it was proposed in Ref.~\refcite{Belotsky:2006fa}.


After they fall down terrestrial surface the in-falling $OHe$
particles are effectively slowed down due to elastic collisions with
matter. Then they drift, sinking down towards the center of the
Earth with velocity \beq V = \frac{g}{n \sigma v} \approx 80 S_3
A_{med}^{1/2} \cm/\s. \label{dif}\eeq Here $A_{med} \sim 30$ is the average
atomic weight in terrestrial surface matter, $n=2.4 \cdot 10^{24}/A_{med}$
is the number of terrestrial atomic nuclei, $\sigma v$ is the rate
of nuclear collisions and $g=980~ \cm/\s^2$.

Then the O-helium abundance the Earth is determined by
the equilibrium between the in-falling and down-drifting fluxes.

The in-falling O-helium flux from dark matter halo is
$$
  F=\frac{n_{0}}{8\pi}\cdot |\overline{V_{h}}+\overline{V_{E}}|,
$$
where $V_{h}$-speed of Solar System (220 km/s), $V_{E}$-speed of
Earth (29.5 km/s) and $n_{0}=3 \cdot 10^{-4} S_3^{-1} \cm^{-3}$ is the
local density of O-helium dark matter. Here, for simplicity, we don't take into account velocity dispersion and distribution of particles in the incoming flux that can lead to significant effect.

At a depth $L$ below the Earth's surface, the drift timescale is
$t_{dr} \sim L/V$, where $V \sim 400 S_3 \cm/\s$ is given by
Eq.~(\ref{dif}). It means that the change of the incoming flux,
caused by the motion of the Earth along its orbit, should lead at
the depth $L \sim 10^5 \cm$ to the corresponding change in the
equilibrium underground concentration of $OHe$ on the timescale
$t_{dr} \approx 2.5 \cdot 10^2 S_3^{-1}\s$.

In underground detectors, $OHe$ ``atoms'' are slowed down to thermal
energies and give rise to energy transfer $\sim 2.5 \cdot 10^{-4}
\eV A/S_3$, far below the threshold for direct dark matter
detection. It makes this form of dark matter insensitive to the
severe CDMS constraints \cite{Akerib:2005kh}. However, in $OHe$ reactions
with the matter of underground detectors  can lead to observable effects.

The equilibrium concentration, which is established in the matter of
underground detectors, is given by
\begin{equation}
    n_{oE}=\frac{2\pi \cdot F}{V} = n_{oE}^{(1)}+n_{oE}^{(2)}\cdot sin(\omega (t-t_0)),
    \label{noE}
\end{equation}
where $\omega = 2\pi/T$, $T=1yr$ and
$t_0$ is the phase. The averaged concentration is given by
\begin{equation}
    n_{oE}^{(1)}=\frac{n_o}{320S_3 A_{med}^{1/2}} V_{h}
\end{equation}
and the annual modulation of concentration is characterized by
\begin{equation}
    n_{oE}^{(2)}= \frac{n_o}{640S_3 A_{med}^{1/2}} V_E
\end{equation}
The rate of nuclear reactions of OHe with nuclei is proportional to the local concentration and
the energy release in these reactions should lead to observable signal.
There are two parts of the signal: the one determined by the constant part and annual
modulation, which is concerned by the strategy of dark matter search
in DAMA experiment \cite{Bernabei:2008yi}.

\subsection{Low energy bound state of O-helium with nuclei}

Our explanation \cite{Bled09,unesco} is based on the idea that OHe,
slowed down in the matter of DAMA/NaI or DAMA/LIBRA detector, can form a few keV bound state with
nucleus, in which OHe is situated \textbf{beyond} the nucleus.
Therefore the positive result of this experiment is explained by reaction
\begin{equation}
A+(^4He^{++}X^{--}) \rightarrow [A(^4He^{++}X^{--})]+\gamma
\label{HeEAZ}
\end{equation}
with sodium and/or iodine.
In detectors with different chemical content such level may not exist at all, or has other value of energy. The rate of reaction (\ref{HeEAZ}) is proportional to temperature and suppressed in cryogenic detectors, making the comparison of their results with DAMA a nontrivial task.

The approach of Refs.~\refcite{Bled09}
and \refcite{unesco} assumes the following picture: at the distances larger, than its size,
OHe is neutral and it feels only Yukawa exponential tail of nuclear attraction,
due to scalar-isoscalar nuclear potential. It should be noted that scalar-isoscalar
nature of He nucleus excludes its nuclear interaction due to $\pi$ or $\rho$ meson exchange,
so that the main role in its nuclear interaction outside the nucleus plays $\sigma$ meson exchange,
on which nuclear physics data are not very definite. When the distance from the surface of nucleus becomes
smaller than the size of OHe, the mutual attraction of nucleus and OHe is changed by dipole Coulomb repulsion. Inside the nucleus strong nuclear attraction takes place. In the result the spherically symmetric potential appears, given by
\begin{equation}
U=-\frac{A_{He} A g^2 exp(-\mu r)}{r} + \frac{Z_{He} Z e^2 r_o \cdot F(r)}{r^2}.
\label{epot}
\end{equation}
Here $A_{He}=4$, $Z_{He}=2$ are atomic weight and charge of helium, $A$ and $Z$ are respectively atomic weight and charge of nucleus, $\mu$ and $g^2$ are the mass and coupling of scalar-isoscalar meson - mediator of nuclear attraction, $r_o$ is the size of OHe and $F(r)$ is its electromagnetic formfactor, which strongly suppresses the strength of dipole electromagnetic interaction outside the OHe "atom".

\begin{figure}[pb]
\centerline{\psfig{file=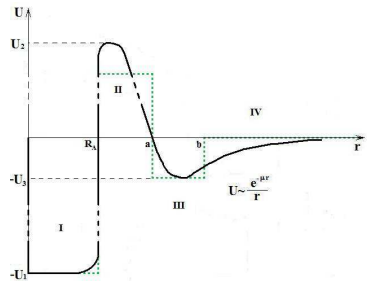,width=2in}}
\vspace*{8pt}
        \caption{The approximation of rectangular well for potential of OHe-nucleus system.}\label{pic23}
\end{figure}

To simplify the solution of Schrodinger equation the potential (\ref{epot}) was approximated in \cite{Bled09}
by a rectangular potential, presented on Fig. \ref{pic23}.

Solutions of Schrodinger
equation for each of the four regions, indicated on Fig. \ref{pic23}, are given in textbooks (see e.g.\cite{LL3}) and
 their sewing determines the condition, under which a low-energy  OHe-nucleus bound state appears in the region III.

\begin{figure}[pb]
\centerline{\psfig{file=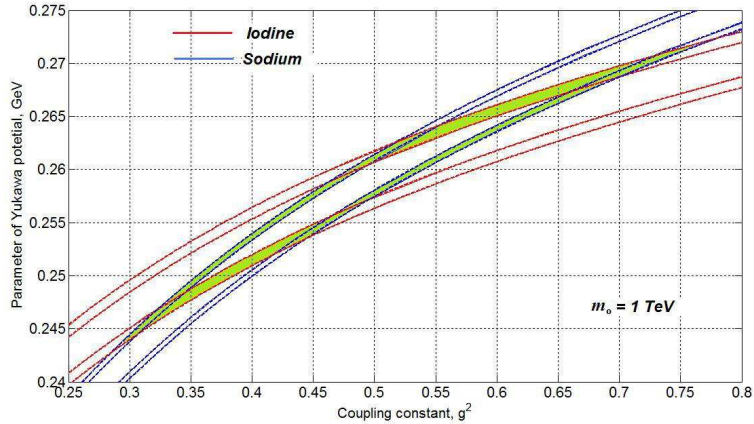,width=4in}}
\vspace*{8pt}
\caption{The region of parameters $\mu$ and $g^2$, for which Na and I have a level in the interval 2-6 keV. For each nucleus two narrow strips determine the region of parameters, at which the bound system of this element with OHe has a level in 2-6 keV energy range. The outer line of strip corresponds to the level of 6 keV and the internal line to the level of 2 keV. The region of intersection of strips correspond to existence of 2-6 keV levels in both OHe-Na and OHe-I systems, while the piece of strip between strips of other nucleus corresponds to the case, when OHe bound state with this nucleus has 2-6 keV level, while the binding energy of OHe with the other nuclei is less than 2 keV by absolute value.}\label{NaI}
   \end{figure}

The energy of this bound state and its existence strongly depend on the parameters $\mu$ and $g^2$ of nuclear potential (\ref{epot}). On the Fig. \ref{NaI} the region of these parameters, giving 2-6 keV energy level in OHe bound states with sodium and iodine are presented. In these calculations \cite{Bled09} the mass of OHe was taken equal to $m_o=1 TeV$.

The rate of radiative capture of OHe by nuclei can be calculated with the use of the analogy with the radiative capture of neutron by proton with the account for: i) absence of M1 transition that follows from conservation of orbital momentum and ii) suppression of E1 transition in the case of OHe. Since OHe is isoscalar, isovector E1 transition can take place in OHe-nucleus system only due to effect of isospin nonconservation, which can be estimated by factor $f \sim 10^{-3}$, corresponding to relative mass difference of neutron and proton. In the result the rate of OHe radiative capture by nucleus with atomic number $A$ and charge $Z$ to the energy level $E$ in the medium with temperature $T$ is given by
\begin{equation}
    \sigma v=\frac{f \pi \alpha}{m_p^2} \frac{3}{\sqrt{2}} (\frac{Z}{A})^2 \frac{T}{\sqrt{Am_pE}}.
    \label{radcap}
\end{equation}

Formation of OHe-nucleus bound system leads to energy release of its binding energy, detected as ionization signal. In the context of our approach the existence of annual modulations of this signal in the range 2-6 keV and absence of such effect at energies above 6 keV means that binding energy of Na-OHe and I-OHe systems in DAMA experiment should not exceed 6 keV, being in the range 2-4 keV for at least one of these elements. The amplitude of annual modulation of ionization signal (measured in counts per day per kg, cpd/kg) is given by
\begin{equation}
\zeta=\frac{3\pi \alpha \cdot n_o N_A V_E t Q}{640\sqrt{2} A_{med}^{1/2} (A_I+A_{Na})} \frac{f}{S_3 m_p^2} (\frac{Z_i}{A_i})^2 \frac{T}{\sqrt{A_i m_p E_i}}=
4.3\cdot10^{10}\frac{f}{S_3^2} (\frac{Z_i}{A_i})^2 \frac{T}{\sqrt{A_i m_p E_i}}.
\end{equation}
Here $N_A$ is Avogadro number, $i$ denotes Na or I,  $Q=10^3$ (corresponding to 1kg of the matter of detector), $t=86400 \s$, $E_i$ is the binding energy of Na-OHe (I-OHe) system and $n_{0}=3 \cdot 10^{-4} S_3^{-1} \cm^{-3}$ is the
local density of O-helium dark matter. The value of $\zeta$ should be compared with the integrated over energy bins signals in DAMA/NaI and DAMA/LIBRA experiments and the result of these experiments can be reproduced e.g. for $E_{Na} = 3 \keV$ and $E_{I} = 5 \keV$.

At the corresponding values of $\mu$ and $g^2$ energy of OHe binding with other nuclei can strongly differ from 2-6 keV. In particular, energy release at the formation of OHe bound state with thallium can be larger than 6 keV. However, for the cross section of radiative capture of thallium by OHe, given by Eq. (\ref{radcap}) and taking into account that thallium content in DAMA detector is 3 orders of magnitude smaller, than NaI, such signal is below the experimental errors.

It should be noted that the results of DAMA experiment exhibit also absence of annual modulations at the energy of MeV-tens MeV. Energy release in this range should take place, if OHe-nucleus system comes to the deep level inside the nucleus (in the region I of Fig. \ref{pic23}). This transition implies tunneling through dipole Coulomb barrier and is suppressed below the experimental limits. The actual rate of these transitions is under our current study.

Since OHe capture rate is proportional to the temperature, it is suppressed in cryogenic detectors by a factor of order $10^{-4}$. The predicted effects of OHe radiative capture in different cryogenic detectors at $T=10 \mK$, $f=10^{-3}$ and $S_3=1$ are given in  Table~\ref{ta1}.
\begin{table}[ph]
\tbl{Effects of OHe in cryogenic detectors.}
{\begin{tabular}{@{}cccc@{}} \toprule
Content & Binding energy & Capture rate &
Counts per day \\
& (keV) & ($10^{-33}\cm^3/\s$) & ($10^{-7} cpd/kg$)\\ \colrule
Ge\hphantom{00} & \hphantom{0}8.6 & \hphantom{0}4.9 & 7.2 \\
Xe\hphantom{00} & \hphantom{0}6.1 & \hphantom{0}4.4 & 3.5 \\
Ar\hphantom{00} & \hphantom{0}15.6 & \hphantom{0}4.8 & 12.8 \\
C\hphantom{000} & \hphantom{0}398.5 & \hphantom{0}2.0 & 17.8 \\ \botrule
\end{tabular} \label{ta1}}
\end{table}

\section{Conclusions}

To conclude, the existence of heavy stable charged particles may not only be compatible with the experimental constraints but even lead to composite dark matter scenario of nuclear interacting Warmer than Cold Dark Matter. This new form of dark matter can provide explanation of excess of positron annihilation line radiation, observed by INTEGRAL in the galactic bulge. The search for stable -2 charge component of cosmic rays is challenging for PAMELA and AMS02 experiments. Decays of heavy charged constituents of composite dark matter can provide explanation for anomalies in spectra of cosmic high energy positrons and electrons, observed by PAMELA, FERMI and ATIC. In the context of our approach search for heavy stable charged quarks and leptons at LHC acquires the significance of experimental probe for components of cosmological composite dark matter.

The results of dark matter search in experiments
DAMA/NaI and DAMA/LIBRA can be explained in the framework of
our scenario without contradiction with negative
results of other groups. Our approach contains distinct
features, by which the present explanation can be distinguished from
other recent approaches to this problem \cite{Edward} (see also
review and more references in Ref.~\refcite{Gelmini}).

The proposed explanation is based on the mechanism of low energy binding of OHe with nuclei.
Within the uncertainty of nuclear physics parameters there exists a range at which OHe
binding energy with sodium and/or iodine is in the interval 2-6 keV. Radiative capture of OHe to this bound state leads to the corresponding energy release observed as an ionization signal
in DAMA detector.

OHe concentration in the matter of underground detectors is determined by the equilibrium between the incoming cosmic flux of OHe and diffusion towards the center of Earth. It is rapidly adjusted and follows the
change in this flux with the relaxation time of few
minutes. Therefore the rate of radiative capture of OHe should experience annual modulations reflected in annual modulations of the ionization signal from these reactions.


An inevitable consequence of the proposed explanation is appearance
in the matter of DAMA/NaI or DAMA/LIBRA detector anomalous
superheavy isotopes of sodium and/or iodine,
having the mass roughly by $m_o$ larger, than ordinary isotopes of
these elements.

Our results show that the ionization signal, detected by DAMA, is proportional to the temperature and should be suppressed in cryogenic detectors. Therefore test of results of DAMA/NaI and DAMA/LIBRA experiments by other experimental groups can become a very nontrivial task, especially, in view of their rejection of electromagnetic part of counting rate in the absence of nuclear recoil.

The presented approach sheds new light on the physical nature of dark matter. Specific properties of composite dark matter and its constituents are challenging for their experimental search. OHe interaction with matter is an important aspect of these studies. In this context positive result of DAMA/NaI and DAMA/LIBRA experiments may be a signature for exciting phenomena of O-helium nuclear physics.

\section*{Acknowledgments}

We would like to thank Pierluigi Belli, Rita Bernabei, Jean Pierre Gazeau and Bernard Sadoulet for discussions and important comments. We express our gratitude to organizers of IWARA09 for cooperation.

\end{document}